\documentstyle[12pt]{article}
\begin{document}
\makeatletter
\parindent 1 PC
\oddsidemargin   -.1 in
\evensidemargin   -.1 in
\topmargin  1.0 cm
\leftmargin -.2 in
\textheight 22 cm
\textwidth  16.5 cm
\setlength{\parsep}{0.5ex plus0.2ex minus0.1ex}

\date{}

\thispagestyle{empty}\setcounter{page}{1}

\title{\bf  Physics of Factorization}
\author{M. Revzen, A. Mann and J. Zak\\
Department of Physics, Technion - Israel Institue of Technology,\\
Haifa 32000, Israel}
\maketitle

{\bf Abstract}\\

 The N {\it distinct} prime numbers that make up a composite number
M allow $2^{N-1}$ bi partioning into two relatively prime factors. Each
such pair defines a pair of conjugate representations. These pairs of 
conjugate representations, each of which spans the M dimensional space are 
the familiar
complete sets of Zak transforms (J. Zak, Phys. Rev. Let.{\bf 19}, 1385 (1967) ) which are the 
most natural representations for periodic systems. Here we show their 
relevance 
to factorizations. An example is provided for the manifestation of the 
factorization.
\bigskip

PACS Numbers 03.65.-w

\newpage  
Shor's discovery \cite{shor} of an algorithm for factorization with quantum 
computers
is considered one of the benchmarks in the developement of quantum 
information theory. In the following we hope to initiate a study of the 
physics of factorization, i.e. the interrelation between factorization and 
physical representations. Our study shows that a finite dimensional 
kq representation (i.e. the Zak transform) \cite{zak} can be viewed as a 
member in a set of pairs of conjugate representations each spanning the  
same finite dimensional phase space. 
Members of the set (Fourier transform is counted as one 
such member) relate to the factorization of the number M which is the 
space dimensionality. Thus we have a correspondence between the set of 
Zak transforms in $M$ dimensional space and the factorization of $M$.
To achieve self containment we shall review some of 
our earlier results \cite{mann}, which are based on   
a general theory of quantum mechanics in finite phase plane \cite{zak2} that 
was developed by
Schwinger \cite{schwinger}. In Schwinger's study the finiteness of  phase 
space is  realized  by applying boundary conditions to  the
coordinate $x$ of the considered wave functions $\psi(x)$, {\it and} to their
Fourier transform $F(p)$:
\begin{equation}\label{period}
\psi(x+Mc)=\psi(x); \ F(p+\frac{2\pi}{c})= F(p)  ,
\end{equation}
where $M$ is an integer and $c$ is constant. (We refer to $c$ as the
scaling constant.) As a consequence of these
boundary conditions, the coordinate $x$ and the momentum $p$ are quantized
($\hbar = 1$ ) and they assume the following discrete values
\begin{equation}
x=sc, ;\ s=1, \dots, M; \ p={2\pi \over Mc}t, \ t=1,\dots, M. 
\end{equation}
Thus Eq.(\ref {period}) implies (\cite{schwinger}) an $M$ dimensional
vector space. 
In our study the operators $\hat x$ and $\hat p$ are replaced by the 
exponential operators $\tau(c)$ and $T(c)$ defined by,
\begin{equation}\label{c}
\tau(Mc)=e^{i x \frac{2\pi}{Mc}}, \; \;\;\;\; \;
T(c)=\;e^{ipc}. 
\end{equation}

We now assume that M is factorizable to 
\begin{equation}\label{factorization}
M=M_{1}^{n}M_{2}^{m}\dots M_{N-1}^{r}M_{N}^{s},  
\end{equation}
with the $M_{i} \ne M_{j}$, $(i \ne j)$  
prime numbers, while $n,m,...,r,s$ are positive integers, i.e., $M$ is made  
of $N$ distinct prime numbers. We now consider the partitioning of this 
product into two factors,   

\begin{equation}\label{kq}
M = M_{a}M_{\tilde a}.
\end{equation}
Here $M_{a}$ incorporates one part of the above $N$ factors while 
$M_{\tilde a}$
contains the other part. We note that this assures that the two numbers
$M_{a}$ and $M_{\tilde a}$ are relative prime, i.e. the equation
\begin{equation} \label{prime}
sM_{a}\;-\;tM_{\tilde a}\;=\;0\;\;[Mod\;M]
\end{equation}
has only the trivial solution for the integers $[s,\;t]$, viz 
$t=M_{a},\;s=M_{\tilde a}$. We further define the two lengths, 
\begin{equation}\label{ab}
a=M_{a}c \;,\;\;\; \ {\tilde a}=M_{\tilde a}c \ .
\end{equation}
An example of one such partitioning for the $M$ given in 
Eq.(\ref{factorization}) is
\begin{equation} \label{prime}
 M_{a}= M_{1}^{n}\;\;{\rm and}\;\; M_{b}= M_{2}^{m} \dots M_{N}^{s}.
\end{equation} 
Next we define two $kq$-representations  based on the two complete 
sets of commuting operators \cite{zak} 
\begin{eqnarray}
\tau(a)\;&=&\;e^{i{\hat x}{2\pi \over a}},\;\;\; T(a)=e^{i{\hat p}a} \label{a}\\
\tau({\tilde a})\;&=&\;e^{i x{2\pi \over \tilde a}},\;\;\; 
T({\tilde a})=e^{ip{\tilde a}} \label{a'}.
\end{eqnarray}
We shall refer to the first pair, associated with
Eq. (\ref{a}), as the ``a'' set and to the second, Eq. (\ref{a'}), as the 
``${\tilde a}$''  set.(Thus, e.g., 
the ``a'' set has its $q$ coordinate: $q=gc,\;\;g=1,\dots,M_a$, while 
the set ``${\tilde a}$'' has its $Q$ coordinates 
$Q=g'c\;\;g'=1,\dots,M_{\tilde a}$.) We have 
\begin{equation}
\left[\tau(a), T(a)\right] = \left[\tau({\tilde a}),T({\tilde a})\right] = 0, 
\end{equation}

but
\begin{eqnarray}\label{kqKQ}
T(a)\;\tau({\tilde a})\;&=&\;\tau({\tilde a})\;T(a)\;\exp(i2\pi 
{M_a\over M_{\tilde a}}),\nonumber\\
T({\tilde a})\;\tau(a)\;&=&\;\tau(a)\;T({\tilde a})\;\exp(i2\pi 
{M_{\tilde a}\over M_a}).
\end{eqnarray}
It therefore follows that the operators $T(a)$ and $\tau(a)$
and their powers form a set of $M$ commuting operators. The same holds for
the operators $T({\tilde a})$ and $\tau({\tilde a})$. This means that the
operators in Eqs. (\ref{a},\ref{a'}) together with their products 
lead to $M^2$ distinct
operators which replace the $M^2$ operators in Eq.(\ref{c}). (Note: a 
Zak transform
may be defined for arbitrary ``a'''s (unrelated to ${\tilde a}$) but only 
when the two factors
$M_{a}$ and $M_{\tilde a}$ of Eq. (\ref{kq}) are relatively prime the two sets 
form a conjugate pair as is discussed below. We study this case only.)\\ 

We now consider the complete sets of eigenvectors of each of the two 
commuting sets of
operators in Eq.(\ref{a},\ref{a'}):
\begin{eqnarray}
\tau(a)|k,q\rangle\;&=&\; e^{iq{2\pi\over a}}|k,q\rangle;\;\; 
T(a)|k,q\rangle = e^{ika}|k,q\rangle, \label{ev1} \\
\tau({\tilde a})|K,Q\rangle\;&=&\; e^{iQ{2\pi\over {\tilde a}}} |K,Q\rangle; \;\;
T({\tilde a})|K,Q\rangle = e^{iK{\tilde a}}|K,Q\rangle.\label{ev2}
\end{eqnarray}
Here $|k,q\rangle$ and $|K,Q\rangle$ are, respectively, the eigenvectors of
the pairs of commuting operators $\tau(a)$, $T(a)$ and
$\tau({\tilde a})$, $T({\tilde a})$ in Eqs.(\ref{a},\ref{a'}).  
In Eq.(\ref{ev1},\ref{ev2}) the variables $k,q,\;K$ and $Q$ assume the
following values \cite{zak},
\begin{eqnarray}
k={2\pi\over Mc}f, \ f&=&1,\dots,M_{\tilde a}, ;\; q=gc, \ g=1,\dots,M_a \label{k}\\
K={2\pi\over Mc}{\bar f},\;{\bar f}&=&1,\dots,M_a,; \; Q={\bar g}c, \
{\bar g}=1,\dots,M_{\tilde a}. \label{K} 
\end{eqnarray}
Note that for a given $M$, Eq. (\ref{factorization}), $M_a$ implies 
$M_{\tilde a}$ so we may distinguish distinct Zak transforms by an extra
label for which we, conveniently, choose the letter ``a''. Thus e.g., the 
Zak transform of Eq. (\ref{ev1}) is written as $|k,q;a\rangle$ while its mate,
 Eq. (\ref{ev2}), is written as $|K,Q;{\tilde a} \rangle$. This extra label is 
somewhat analogous to the characterization of irreducible representations via 
the values of the appripriate Casimir operator, \cite{hamermesh}, in as much 
as it does not relate to an eigenvlues but to the representation as a whole.
  
We now evaluate  $\langle a; k,q|K,Q;a\rangle$. To this end we use, 
\cite{zak}, that the x- represenatoion of the eigenfunctions of the ``a'' 
set operators
 (cf. Eq. (\ref{ev1}) is given by,
\begin{equation}\label{x}
\langle x|k,q;a \rangle\;=\;{1 \over M_{\tilde a}}\sum_{s=1}^{M_{\tilde a}}
e^{iksa}\Delta(x-q-sa),
\end{equation}
with a similar equation for the ``$|K,Q;{\tilde a}\rangle$'' set. 
$\Delta(x)$ is 
1 when $x$ is a 
multiple of $Mc$, and it vanishes otherwise. (We note in passing that
the set $|x\rangle$ may be viewed as an ``a'' set with $M_{a}=M$, and 
$M_{\tilde a}=1$ i.e. it may be written as $|k={2\pi\over a},q;a=M\rangle$,
 since
this coincides with the complete set associated with the commuting 
operators, (cf. Eq. (\ref{ev1}))  $\tau(Mc)$ and $T(Mc)$, the latter  being 
unity in our $M$ dimensional space while ``$q$'' spans the whole space. Its 
mate involves the eigenfunctions of 
the momentum operator, and, equivalently, forms the eigenfunctions of $\tau(c)$
and $T(c)$, Eq. (\ref{ev2}). This pair of conjugate represenations are then
 the 
``familar'' Fourier represenations as is further clarified below.) With 
these  we may write (the extra label, ``a'', for the states is understood, 
and will be put explicitly only if required), 
\begin{eqnarray}
\langle k,q|K,Q\rangle&=&\sum_{x=c}^{Mc}\langle kq|x\rangle \langle x|KQ
\rangle\\                    &=&{1 \over \sqrt{M_{a}M_{\tilde a}}}
\sum_{s=1}^{M_{\tilde a}}\sum_{t=1}
^{M_{a}}\exp(-iksa+iKt{\tilde a})\Delta(Q+t{\tilde a}-q-sa)
\end{eqnarray}
recalling that $a=M_{a}c$ and ${\tilde a}=M_{\tilde a}c$ 
 (Eq.(\ref{ab})), the above expression does not vanish only for
\begin{equation}\label{st}
 tM_{\tilde a}-sM_{a}= \frac{q-Q}{c}\;\equiv \;r\;\; [{\rm Mod}\; M].
\end{equation}     
For $M_{a}$ and $M_{\tilde a}$ relatively prime (Eq. (\ref{prime})),
then the Eq. (\ref{st}) has a unique pair $[s,t]$ for each $r\;=\;1,\dots M$.
In this case \cite{mann,schwinger},
\begin{equation}\label{conjugate}
\langle k,q|K,Q\rangle = 
\frac{1}{\sqrt{M_aM_{\tilde a}}} \exp (-iksa + iKt{\tilde a})
\Delta(Q + t{\tilde a} - q -sa),
\end{equation}
where $\Delta(x)$ does {\it not} vanish only when $x= 0\;\; ({\rm Mod}\; Mc)$. 
(Note: for the special case with $a=Mc$ (hence ${\tilde a}=c$) 
Eq. (\ref{conjugate})
gives for the right hand side since, in this case, the set $|k,q\rangle$
is independent of k: in this finite dimensional phase space the 
operator $\exp(ipMc)=1$. Similarly, for its mate, the set $|K,Q\rangle$ is
independent of Q: the operator $\exp(ix{2\pi \over c}$ is
unity for all Q. Thus 
$${1 \over \sqrt {M}}\exp (iKt)\Delta(Q-q+t)\;=\;{1 \over \sqrt {M}}\exp(iKq),$$
i.e. this is the ``familiar'' Fourier representation (we have not included an 
irrelevant overall phase factor)).\\ 
 We note that  
\begin{equation}\label{N}
2^{(N-1)}\;=\;{\rm number\;of\;bi\;partitioning\;of\;Eq.\;(\ref{factorization})} 
\end{equation}
of an arbitrary $M$ (given by Eq. (\ref{factorization}))
to such conjugate pairs (where $N$ is the number 
of {\it distinct} prime factors making the number $M$. This follows from 
the fact that $2^{N}$ is the number of bi partitions of 
N distinct items whose relative ordering is immaterial and that we need only 
half of these (since we are free to label either as an ``a'' or ``$\tilde a$'' set). We note that whenever a factor, say $M_{j}^{n}$ occurs (in $M_{a}\; 
{\rm or}\;M_{\tilde a}$) it 
is considered only once as representing the prime number $M_{j}$.
Thus our label ``a'' for the sets of conjugate pairs has N distinct values. 
We reiterate that Eq. (\ref{conjugate}) implies that 
$|\langle a;k,q|K,Q;{\tilde a}\rangle|$ does not depend on $q$ and $Q$ nor on $k$ and 
$K$. One has to keep
in mind, however, that $s$ and $t$ in the phase in Eq.(\ref{conjugate}) are 
determined
by $r$ in Eq.(\ref{st}). 
The result Eq. (\ref{conjugate}) shows that when the system is in the 
eigenstate $|K,Q;a\rangle$
of the commuting operators $T({\tilde a})$ and $\tau({\tilde a})$ 
[see Eq.(\ref{ev1}),\ref{ev2})],
the probability of measuring $k$ and $q$ does not depend on $k$ and $q$. The
same can be said about measuring $K$ and $Q$ in the eigenstate
$|k,q;a\rangle$. We can therefore claim that the two sets of commuting
operators in Eq.(7) are conjugate \cite{mann,englert}.
An important property of conjugate operators is as follows \cite{mann}.
 When the ``a'' set operators in
Eq.(7) operate on the eigenvectors of the ``$\tilde a$'' set, the eigenvalues 
of these
eigenvectors are shifted. Let us first find the eigenvalues of the vectors
$T(a)|K,Q;{\tilde a}\rangle$. We have, by using the first equation in 
Eq. (\ref{kqKQ}), and Eq.(9)
$$
\tau(\tilde a) T(a)|K,Q;{\tilde a}\rangle = e^{i(Q-a){2\pi \over {\tilde a}}}
T(a)|K,Q;{\tilde a}\rangle .
$$
Thus the operators of the ``a'' set, when acting on an eigenstate of its 
conjugate mate span the whole space in a unique way, with a similar
effect for operators of the ``$\tilde a$'' set acting on state of the ``a'' set. Thus
each conjugate pair of the set spans the whole $M$ dimensional space. The 
total number of such pairs equals $2^{(N-1)}$, where is N the number of 
distinct prime numbers that make up $M$. 
We now consider a ``physical'' implication of this factorization scheme. 
A simple illustration of it is gained via the familiar Fourier 
transform which, from the present vantage point is viewed as ``factorizing'' 
the dimensionality $M$ as  $M = M\cdot1$, i.e. in the ``a'' set we have  $M_a = M$ (and, hence, $M_{\tilde a}=1$), in other words it is the familiar x 
space (of dimension $M$) and
its mate the ``$\tilde a$'' set, the eigenfunctions of $\tau(c)$ and $T(c)$,
 is here the momentum space (also of dimension 
$M$). We consider for this conjugate pair the completely delocalized  state 
$|\psi\rangle$, i.e.

\begin{equation}
\langle x_{n}|\psi\rangle = {1\over \sqrt M},\;\;n=1,...,M .
\end{equation}
This state, expressed in terms of its conjugate set 
(with $(K_{m}={2\pi\over Mc}m)$), is
\begin{eqnarray}\label{deloc}
\langle K_{m}|\psi \rangle\;&=&\; \sum_{n=1}^{M}\langle K_{m}|x_{n}\rangle
\langle x_{n}|\psi\rangle\;=\;{1\over M}\sum_{n} e^{iK_{m}x_{n}} \nonumber \\
    &=& {1\over M}e^{i(2\pi m/M)}\frac{1-e^{i2\pi m}} {1-e^{i2\pi m/M}}\;=
\;m\delta_{K_{m},K_{M}}.
\end{eqnarray}
Here $\delta_{n,m}$ is the Kronecker delta.(We have not included an irrelevant
overall phase factor.) Thus the state is completely localized in
its conjugate set space (p space), we note that we have retrieved the factor 
$1$ in the 
factorization  $M = M\cdot1$, given above in that the state is localized at a 
point. Thus for the factorization $M=M\cdot1,$ when we have one member of 
the pair of conjugate set one dimensional, complete delocalization in the 
first 
(the ``a''member) leads to complete localization (to one eigenstate) in the
the ``$\tilde a$'' set. We now consider an arbitrary pair whose ``a'' member 
involves
the factor $M_{a}$ whilst its mate, ``$\tilde a$'', involves the factor
 $M_{\tilde a}$, with,
of course, $M = M_{a}M_{\tilde a}$. We take
$M_{\tilde a} > M_{a}.$ We now consider a completely delocalized state in the 
``a'' set, e.g. a particle spread uniformaly over the coordinates of this 
member of the set, i.e.,
\begin{equation}\label{delocalized}
\langle a;k,q|\psi \rangle\; = \; { 1 \over {\sqrt M}},
\end{equation}
and wish to evaluate its value in the ``$\tilde a$'' coordinates, i.e. we 
seek the value of 
$\langle a;K,Q|\psi \rangle$ (again deleting the ``a'' label as implicit in the following),
\begin{equation}
\langle K,Q|\psi \rangle\;=\;\sum_{f,g}^{M_{b},M_{a}}\langle K,Q|k,q\rangle
\langle k,q|\psi \rangle,
\end{equation}
where
\begin{equation}
 k\;=\;{2\pi\over Mc}f,\;\;f\;=\;1,2 \dots M_{\tilde a},\;q\;=\;gc,\;\;g\;=\;
1,...M_{a}.
\end{equation}
Substituting Eq. (\ref{conjugate}) and performing the 
summation over the index $f$ we get, in complete analogy to Eq. (\ref
{deloc}) (we do not give an irrelevant, overall phase factor.), that
 only for
 $s\;=\;M_{\tilde a}$ we do get a contribution, which 
is $M_{\tilde a}$. Now
since $Q$ is fixed, only one $t$ contributes \cite{r}, viz $t=M_{a}$. This is 
seen as follows: $|q - Q| \le M_{\tilde a}c$, as $c \;\le \;q\;\le\;M_{a}c$ 
and,
$c\;\le\;Q\;\le\;M_{\tilde a}c$, with $M_{a}\;<\;M_{\tilde a}$. Hence 
Eq.(\ref{st}) with $s\;=\;M_{\tilde a}$ can only be satisfied with $q\;-\;Q\;
=\;0,$
 hence $t\;=\;M_{a}$. We have then that for each  $Q\;=\;gc, g\;=\;1,..M_{a}$ 
 and  $K\;=\;{2\pi \over Mc}f, f\;=\;1,..,M_{a}$ 
\begin{equation}\label{overlap}
\langle K,Q|\psi\rangle\;=\;{1 \over M_{a}},
\end{equation}    
and it vanishes elsewhere (an overall phase factor was not included). This 
result assures the correct normalization of 
$|KQ\rangle$, since it implies probability of ${1 \over M_{a}^{2}}$ and
there are $M^{2}_{a}$ non vanishing terms. Thus this result generalizes the 
result 
above for the ``usual'' Fourier decomposition to give that a completely 
delocalized state in the ``a'' set ($|k,q>$ representation) is a state 
 which, in the ``$\tilde a$'' set ($|K,Q\rangle$ representation is localized 
over the square (in phase space) covering $M_{a}^{2}$ spots where $M_{a}$
is one of the appropriate factors of  $M$.
 
To summarize, we showed that given a number M that can be factorized into N
distinct prime numbers (one that occurs more than once is counted only once)
and hence may be bipartitioned into $2^{N-1}$ products of the form 
$M=M_{a}M_{\tilde a},$
 allows the definiton of $2^{N-1}$ Zak transform conjugate pairs. The first 
such partioning, $M_{a}=1,\; M_{\tilde a}=M$ leads to the familiar 
Fourier transform pairs while the next N bipartitioning with $M_{a}$ being 
one of counted N primes leads to the appropriate Zak transforms conjugate
pairs each with one member having the dimension of the factor. Other 
partitions yield the other members of Zak transform conjugate pairs. To each
such conjugate pair a ``physical'' manifestation of the factor such as 
$M_{a}$ is gained by noting that the localization dimension of a state 
of one conjugate pair member of $M_{a}^{2}$ is associated with completely 
delocalization of that state in its conjugate mate representation.\\
  
We wish to comment briefly on the choice of the scaling factor c (Eq. (1)).
Physically the dimensionality of the phase space was determined by two 
conditions. The first involves the periodicity in space ,$\psi(x+L)=\psi(x)$,
the second involves the periodicity in the Fourier transposed space,
$F(p+{2\pi \over c})=F(p)$. The two requirements may be consistent for the 
same physical system if $L=Mc$ for some integer $M$. This integer is the 
dimensionality of the vector space and the number we factorized in our 
study. The number of conjugate pairs of Zak transforms depends only on the 
number of distinct primes (cf. Eq. (\ref{factorization})), viz N ,
Eq. (\ref{N}). Thus one may consider varying $c$, and hence $M,$ in such a way 
as to preserve the number of primes and thereby retaining the same class of
Zak transform pairs that retain the same spatial periodicity. This may be 
achieved as follows. Rewrite Eq. (\ref{factorization}) as 
\begin{equation}
Mc\;=\;M_{1},\dots, M_{N}M_{1}^{n-1},\dots, M_{N}^{s-1}c\;=\;{\bar M}{\bar c}.
\end{equation}
With ${\bar M}=M_{1},\dots, M_{N}$ and ${\bar c}=M_{1}^{n-1},\dots, 
M_{N}^{s-1}c$.
Now the phase space dimensionality is determined by requiring the 
Fourier transform be periodic in ${2\pi\over {\bar c}}$. With this, the 
bi factorization to two relatively prime factors each with  
its distinct  Zak transform label, is into primes of first order only, so we 
may think of ${\bar c}$ as leading to an irreducible set of Zak 
transforms.\\      
 Further insight into our results and into the conjugacy in general may be 
gained 
by an alternative approach to the problem. Thus we consider two distinct Zak
transforms, the first charaterized by a length ``a'' and defined through the 
two basic commuting operators ($\hbar$=1) \cite{zak3},
$$
exp(i x{2\pi \over a}),\;\;\;exp(ipa),
$$
while 
the second is similarly characterized by a length ``b''. We now consider
a scaling factor c such that 
$$a\;=\;M_{a}c,\;\;{\rm and}\;\;b\;=\;M_{b}c.$$
with $M_{a},\;M_{b}$ integers. To relate to our problem of {\it finite} phase 
space we consider the length $Mc$ with,
$$
M\;\;=\;\;M_{a}M_{b}.
$$
For  $M_{a}$ and $M_{b}$ relatively prime (i.e. $M_{b}\equiv M_{\tilde a}$) we 
have that the equation
$$sM_{a}\;+\;tM_{b}\;=\;0\; [{\rm Mod}\;M],$$
has a unique solution: $sM_{a}\;=\;0 \;[{\rm Mod}\;M]$, $tM_{b}\;=\;0\;
[{\rm Mod}\;M].$ In these cases  the two Zak transforms
are conjugate in the $M$ dimensional space thus defined, and, the 
factorization of $M$ to $N$ distinct primes, Eq.(\ref{factorization}), implies 
 that there are $2^{N-1}$ distinct relative prime pairs $M_{a},M_{b}$. Thus
 we are led to $M$ which is factorized by the $2^{N-1}$ sets of conjugate Zak
transform pairs. These include the pair with $M_{a}=1,\;M_{b}=M$ which is the 
more familiar Fourier tranform in M dimensions. Here as in all cases, we do 
not count separately the reverse ordering, viz $M_{a}=M, M_{b}=1,$ thereby 
 illustrating the source of $-1$ in the number, $2^{N-1}$, of Zak transform 
conjugate pairs that factorizes the dimensionality number $M$.\\

We summarize our main result by noting that a correspondence was established 
between the factorization of a number $M$ in terms of
 $N$ distinct prime numbers and the number of distinct conjugate
Zak transform pairs, each spanning an $M$ dimensional space. Thus if $M$ is 
given in terms of $N$ (distinct) primes (each counted once regardless of 
the number of times it appears in the factorization), $M$ allows $2^{N-1}$
bipartitioning to two relative prime factors. This number, $2^{N-1}$, is the 
number of (distinct) conjugate Zak transform pairs, each spanning the $M$
dimensional space. We have given an example where, in a particular Zak 
transform representation, the constituent factor may be observed: a state 
that is 
completely delocalized in the representation of one member of a pair of
conjugate Zak transform, is localized (evenly) within the dimensionality
that equals the smaller factor that charactarizes the other member of the 
pair.\\


\begin{thebibliography}{99}
\bibitem{englert} The $\Delta$ ( cf. Eq. (10)) in conjuction with $M_{1}$
 and $M_{2}$
being relatively prime implies that  $\langle k,q|K,Q\rangle$ is 
a phase factor. This is studied in a forthcoming book by B-G Englert.
\end{thebibliography}
\end{document}